\documentclass[twocolumn]{cinc}

\usepackage{amsmath, graphicx, xurl, hyperref, amssymb, comment, booktabs, tabularx, multirow, rotating, adjustbox, bbold}
\usepackage[dvipsnames]{xcolor}
\newcolumntype{L}{>{\raggedright\arraybackslash}X}
\newcolumntype{C}{>{\centering\arraybackslash}X}

\begin{document}
\bibliographystyle{cinc}

\title{ArNet-ECG: Deep Learning for the Detection of Atrial Fibrillation from the Raw Electrocardiogram}


\author {Noam Ben-Moshe$^{1,2}$, Shany Biton$^{2}$, Joachim A. Behar$^{2}$ \\
\ \\ 
 $^1$Faculty of Computer Science, Technion-IIT, Haifa, Israel \\
 $^2$Faculty of Biomedical Engineering, Technion-IIT, Haifa, Israel 
}

\maketitle
\begin{abstract}
\textbf{Introduction:} Atrial fibrillation (AF) is the most prevalent heart arrhythmia. AF manifests on the electrocardiogram (ECG) though irregular beat-to-beat time interval variation, the absence of P-wave and the presence of fibrillatory waves (f-wave). We hypothesize that a deep learning (DL) approach trained on the raw ECG will enable robust detection of AF events and the estimation of the AF burden (AFB). We further hypothesize that the performance reached leveraging the raw ECG will be superior to previously developed methods using the beat-to-beat interval variation time series. Consequently, we develop a new DL algorithm, denoted ArNet-ECG, to robustly detect AF events and estimate the AFB from the raw ECG and benchmark this algorithms against previous work. \textbf{Methods:} A dataset including 2,247 adult patients and totaling over 53,753 hours of continuous ECG from the University of Virginia (UVAF) was used.   
\textbf{Results:} ArNet-ECG obtained an F\textsubscript{1} of 0.96 and ArNet2 obtained an F\textsubscript{1} 0.94. \textbf{Discussion and conclusion}: ArNet-ECG outperformed ArNet2 thus demonstrating that using the raw ECG provides added performance over the beat-to-beat interval time series. The main reason found for explaining the higher performance of ArNet-ECG was its high performance on atrial flutter examples versus poor performance on these recordings for ArNet2.
\end{abstract}

\section{Introduction}
Atrial fibrillation (AF) is the most prevalent heart arrhythmia \cite{bjorck2013atrial, haim2015prospective}. Today, there is up to 13\% of individuals with AF are misdiagnosed \cite{quer2020screening}. High gaps in AF management exist including the lack of effective decision support tools for risk prediction and diagnosis. On the ECG, AF is characterized by an irregular beat-to-beat time interval variation (RR-interval) resulting from the fibrillation of the atria which excites the atrioventricular node at a very high rate. This in turns, causes the atrioventricular node to fire in a chaotic and highly irregular manner and which consequently leads to an irregular polarization of the ventricles. Furthermore, it affects the morphology of the sinus waveform in such a way that the ‘traditional’ P wave is replaced with the appearance of low amplitude f-waves. 
Atrial flutter (AFL) is another supraventricular arrhythmia which has some similarities in its clinical management to AF and so we included AFL within the AF group as previously done by others \cite{carrara2015heart}. AFL results from an abnormal circuit inside the right atrium, or upper chamber of the heart. AFL is characterized by regular RR beats and the presence of flutter waves. In this research we develop a new deep learning (DL) algorithm for AF events detection from long continuous raw ECG. We hypothesize that the raw ECG may yield superior performance compared to using the beat-to-beat time series derived from the ECG since the raw ECG holds both morphological information and rhythm information. This research aims to create a DL algorithm, denoted ArNet-ECG, for the detection of AF events from long continuous ECG recordings.

\section{Methods}
\subsection{Dataset}
The University of Virginia dataset (UVAF) was used. The UVAF \cite{carrara2015heart, moss2014local} consists of ECG recordings of patients for whom the University of Virginia health system physicians ordered Holter monitoring from December 2004 to October 2010.  This dataset contains $n=2,247$ annotated recordings of individual patients over the age of 18 years old. Recordings were sampled at 200 Hz. The median and interquartile age for the recordings were 57 (40-71), recording durations 24 (24-24) hours long. UVAF was used to train and evaluate the DL models. In order to automatically assess the quality of the raw ECG files and discard those that were too noisy, we adopted the R-peak quality criterion bSQI \cite{behar2013ecg}. More information about the dataset is described in \cite{chocron2020remote}. Out of the UVAF dataset 100 recordings from unique patients were selected and used as test set. All other recordings were used as training set and validation set. The test set was re-annotated by an intern cardiologist. Specifically, supraventricular arrhythmias were annotated in three categories:  (1) AF; (2) AFL; (3) Other supraventricular tachycardias such as Wolf-Parkinson-White and intranodal tachycardias and (4) other such as NSR that where not labelled as was done in \cite{Biton2022gengeralization}. Furthermore, for the specific purpose of the main experiments presented in this work the annotations for AF and AFL were grouped under the single label denoted as AF\textsubscript{l}.

\subsection{Preprocessing}
Recordings of patients under 18 years old were also excluded. Corrupted recordings were excluded as described in Chocron et al. \cite{chocron2020remote}. This resulted in a total number of 2,237 recordings used for our experiments. Each recording was divided into 30 seconds non-overlapping windows. A total of 6,424,793 windows were available from the UVAF after the exclusion criteria were applied. For each window, we defined its rhythm label as the reference label most represented over this window. 


\subsection{AF\textsubscript{l} groups definition}
Patient were divided into four groups based on their AFB as defined in Chocron et al \cite{chocron2020remote}, i.e Non-AF, AF\textsubscript{mild}, AF\textsubscript{mod} and AF\textsubscript{sev}. The groups were defined as follows: Non-AF: total time spent in AF below 30 seconds. Mild AF (AF\textsubscript{mild}): total time spent in AF above 30 seconds \cite{kirchhof20162016} and AFB below 4\% \cite{boriani2014device}, Moderate AF (AF\textsubscript{mod}): AFB in the range 4-80\%, Severe AF (AF\textsubscript{sev}): AFB above 80\%.

\subsection{ArNet-ECG}
We used a Residual Network (ResNet) architecture \cite{ResNetOrig} as was previously used in the context of arrhythmia classification taking as input the RR time series \cite{Biton2022gengeralization}. The model takes a window of 30 seconds of raw ECG as input and outputs a binary value (AF\textsubscript{l} or non-AF\textsubscript{l}).  ArNet-ECG architecture is a stack of residual 1D-Convolutional (CNN) blocks followed by dense layers. Each block has 1D-CNN layers with a Batch Normalization (BN) layers followed by a Rectified Linear Unit (ReLU) activation and a shortcut connection with Max Pooling. The output of the residual blocks is then passed to three successive dense layers. The last dense layer of size 1 outputs a binary prediction for the label of the window. The loss function used was a weighted binary cross-entropy loss which gave more weight to the AF windows since the classes are highly imbalanced. 
During training we optimized with Adam \cite{kingma2017adam} algorithm and used dropout regularization. One of the advances of residual blocks is the skip connections that enhance the rate of information transferred throughout the network by connecting layers earlier in the network with layers later in the network as was suggested in \cite{ResNetECG2017}. 

\subsection{Performance statistics}
The performance of the models were assessed as in Chocron et al.  \cite{chocron2020remote}. The following statistics were computed to assess the models performance on the individual 30 seconds windows: sensitivity (Se), specificity (Sp), positive predictive value (PPV), the area under the receiver operating characteristic  (AUROC) and the harmonic mean of the precision and recall (F\textsubscript{1}). The threshold on the models output probabilities was defined as the point which maximizes the validation set F\textsubscript{1}.
The AFB and the absolute AFB estimation error (E\textsubscript{AF} (\%)) were defined as in Chocron et al. \cite{chocron2020remote}. The AFB is defined as follows:
$$AFB = \frac{\sum_{i=1}^{N} l_{i} \times \mathbb{I}_{i}}{\sum_{i=1}^{N} l_{i}}$$
with N available windows, $l_{i}$ length of the $i^{th}$ window and $\mathbb{I}_{i}$ the unity operator equal to 1 for AF and zero otherwise. In the raw-ECG case, the length of each window is 30 seconds which corresponds to 6000 samples at a sampling frequency of 200Hz.
The E\textsubscript{AF} (\%) is defined as:
$$E\textsubscript{AF} (\%) = \frac{\sum_{i=1}^{N} l_{i} \times (\hat{y}_{i}-y_{i})}{\sum_{i=1}^{N} l_{i}}$$
where $y_{i}$ is a binary value representing the window label and $\hat{y}_i$ is the predicted binary label by the model when in for both 1 for AF and 0 otherwise. 
ArNet-ECG classifies 30 seconds windows whereas ArNet2 takes as input 60-beat windows. In order to benchmark both algorithms, their outputs were aligned to provide a classification (AF, non-AF) for each 5-sec window. The nonparametric Mann–Whitney rank test was used to determine whether there is a statistical significance if the F\textsubscript{1} between ArNet-ECG and ArNet2. For that purpose the F\textsubscript{1} was 1000 times computed on randomly sampled 80\% of the test set. The procedure was used to obtain the distribution of the F\textsubscript{1} and the Mann–Whitney rank test was applied on these F\textsubscript{1}  distributions.

\section{Results} 
\subsection{Model performances} \label{Model performances}
Best performance on the validation set were achieved after training the model for two epochs. For single window classification ArNet-ECG reached an F\textsubscript{1} of 0.96 and an AUROC of 0.99. The median and the interquartile range (Q1-Q3) of the absolute AFB estimation error was $|$E\textsubscript{AF}(\%)$|$ 0.41\%(0.03\%-2.45\%) on the test set. The $|$E\textsubscript{AF}$| (\%)$ distributions for the four AF severity levels is shown on Figure \ref{fig:af_burden_plot}. Table \ref{table:AFB error on UVAF} summarizes the results of ArNet2 and ArNet-ECG. ArNet-ECG F\textsubscript{1} performance was significantly higher than ArNet2 (P-Value $< 0.001$).

\begin{table}[htb!]
    \centering
    \begin{tabular}{c c c c c c} 
     \hline
      Model & F\textsubscript{1}  & AUROC  &Se & Sp \\ [0.5ex] 
     \hline
    ArNet2 & 0.94  & \underline{0.99}&  0.93 & \underline{0.96} \\ [0.5ex] 
    ArNet-ECG & \underline{0.96} & \underline{0.99} &\underline{0.96} & \underline{0.96} \\ [0.5ex]
     \hline
    \end{tabular}
    \caption{Results for windows classification on the UVAF test set (n=100).}
    \label{table:AFB error on UVAF}
\end{table}

\vspace{-15pt}
\subsection{Error Analysis}
\begin{figure*}[htb!]
    \centering
    \includegraphics[page=1,width=\textwidth]{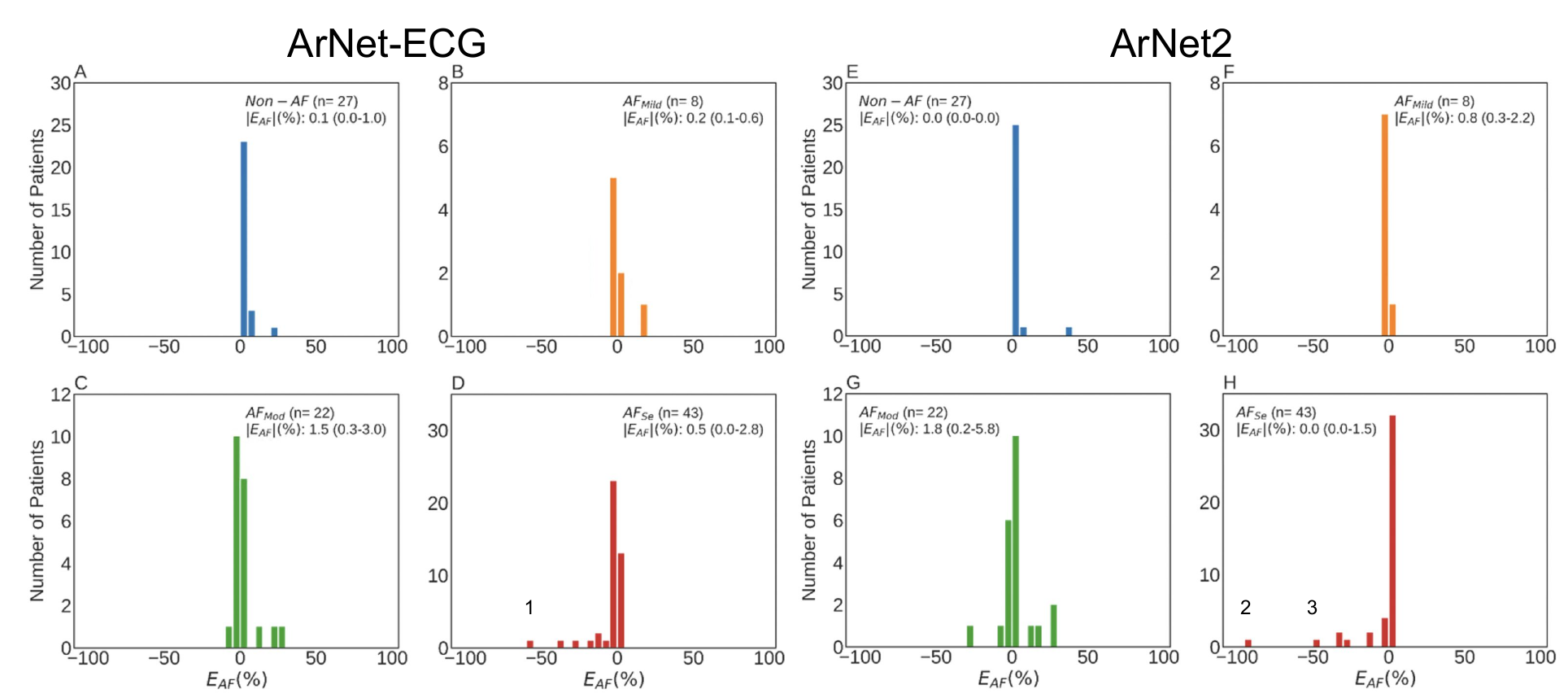}
\caption{Histogram of the median absolute AFB estimation error $|$E\textsubscript{AF}$| (\%)$ for the test set per different AF\textsubscript{l} severity labels: Non-AF (panel A and E), $AF_{mil}$ (Panel B and F), $AF_{mod}$ (panel C and G) and $AF_{sev}$ (panel D and H). On the left results of ArNet-ECG this work)  and on the right results of ArNet2 \protect\cite{Biton2022gengeralization}.}
    \label{fig:af_burden_plot}
\end{figure*}

\begin{figure*}[htb!]
    \centering
    \includegraphics[page=1,width=\textwidth]{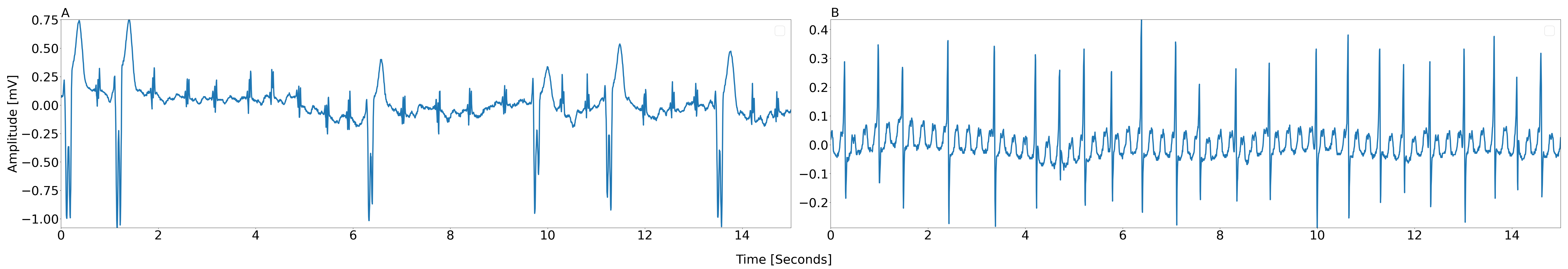} 
    \caption{Error analysis examples. Panel A: example of an ECG window with a reference label of AF as well as containing premature ventricular contraction beats. This window was incorrectly classified by ArNet-ECG but correctly classified by ArNet2. Panel B: example of an ECG window with a reference label of AFL. This section was classified correctly by ArNet-ECG but mis-classified by ArNet2.} 
    \label{fig:error_examples}
\end{figure*}

We observe that ArNet-ECG was able to detect 98.77\% (3060/3098) of the AFL windows. In the test set there were four patients with AFL. We compared results on all windows from these four patients as shown in Table \ref{table:AFL_pat_error_on_UVAF}. Thus ArNet-ECG was able to detect AFL windows significantly better than ArNet2 which often miss-classifies a large number of AFL windows. This is because the beat-to-beat is relatively regular in AFL versus AF. For that reason, it may be challenging to identify AFL events from the RR time series and rather the information on the presence of AFL will mostly be contained in the ECG waveform. 
In Figure \ref{fig:af_burden_plot} the median absolute AFB estimation error $|$E\textsubscript{AF}$| (\%)$ for the test set per different AF severity labels  of ArNet2 and ArNet-ECG is shown. For both ArNet2 and ArNet-ECG there were only errors higher than 45\% in the AF\textsubscript{sev} group. The patient denoted as 1 in Figure \ref{fig:af_burden_plot} got an $|$E\textsubscript{AF}$| (\%)$ of 57.8\% by ArNet-ECG but was accurately classified by ArNet2. This recording contains both AF and ventricular extrasystoles which might have caused the mis-classification by ArNet-ECG. Figure \ref{fig:error_examples} (A) shows an ECG segment misclassified by ArNet-ECG but correctly classified by ArNet2. In Figure \ref{fig:af_burden_plot} the patient denoted as 2 obtained an $|$E\textsubscript{AF}$| (\%)$ of 0.1\% by ArNet-ECG but  $|$E\textsubscript{AF}$| (\%)$ of 90.0\% by ArNet2. This mistake is probably due to the fact that this patient has a low average heart rate at 63 bpm and high beat-to-beat variability which is different than the typical AF cases where both a high variability and high average heart rate are observed. In Figure \ref{fig:af_burden_plot} the patient noted as 3 obtained an $|$E\textsubscript{AF}$| (\%)$ of 0.1\% by ArNet-ECG but  $|$E\textsubscript{AF}$| (\%)$ of 47.3\% by ArNet2. This patient had reference annotations of 100\% AFL. Shown in Figure \ref{fig:error_examples} (B) is an ECG segment classified correctly by ArNet-ECG but wrongly classified by ArNet2. The beat-to-beat variation in this example is rather regular because it is under an AFL event and consequently ArNet2 missed the abnormality. 
\begin{table}[htb!]
    \centering
    \begin{tabular}{c c c c c c} 
     \hline
      Model & F\textsubscript{1}  & AUROC  &Se & Sp \\ [0.5ex] 
     \hline
    ArNet2 & 0.70  & 0.97&  0.54 & \underline{0.99} \\ [0.5ex] 
    ArNet-ECG & \underline{0.98} & \underline{0.99} & \underline{0.97} & \underline{0.99} \\ [0.5ex]     
     \hline
    \end{tabular}
    \caption{Results for windows classification from patients with AFL from the UVAF test set (n=4).}
    \label{table:AFL_pat_error_on_UVAF}
\end{table}

\section{Discussion and Conclusions}
We introduce ArNet-ECG, an algorithm for the detection of AF events from the raw ECG. ArNet-ECG performance reached an F\textsubscript{1} of 0.96 which outperform STOTA performance obtained with ArNet2. Most AFL examples were correctly classified by ArNet-ECG. This can be explained by the fact that raw ECG was used as input to the network and thus even in the case of regular beat-to-beat interval variation in AFL the network learns to recognize flutter waves (f-waves). Future work will include the integration of a recurrent neural network unit to take into account the time dependency between consecutive windows as well as assessing the generalization of ArNet-ECG on external test sets including different ethnic group. 

\balance

\section*{Acknowledgments}  
This research was supported for NBM, SB and JAB by a grant (3-17550) from the Ministry of Science \& Technology, Israel \& Ministry of Europe and Foreign Affairs (MEAE) and the Ministry of Higher Education, Research and Innovation (MESRI) of France. NBM, SB and JAB also acknowledge the support of the Technion-Rambam Initiative in Artificial Intelligence in Medicine.

\bibliography{template}


  
  
      

\vspace{-15pt}
\begin{correspondence}
Dr. Joachim Behar (jbehar@technion.ac.il)
\end{correspondence}

\end{document}